# Bees with attitude: the effect of gusts on flight dynamics

Timothy Jakobi[1,*], Dmitry Kolomenskiy[2], Teruaki Ikeda[3], Simon Watkins[1], Alex Fisher[1], Hao Liu[3], Sridhar Ravi[1]

[1]School of Aerospace Mechanical and Manufacturing Engineering, RMIT University, Australia
[2]Japan Agency for Marine-Earth Science Technology (JAMSTEC), Japan
[3]Graduate School of Engineering, Chiba University, Japan

**Summary statement**

This article examines the effect of Nature's common airborne forces on the flight of a small flapping organism and proposes several control strategies that could be important for all natural and engineered flight at this scale.

**Abstract**

Flight is a complicated task at small scales in part due to the ubiquitous unsteady air which contains it. Flying organisms deal with these difficulties using active and passive control mechanisms to steer their body motion. Body attitudes of flapping organisms are linked with their resultant flight trajectories and performance, yet little is understood about how discrete unsteady aerodynamic phenomena affect the interlaced dynamics of such systems. In this study, we examined freely flying bumblebees subject to a single discrete gust to emulate aerodynamic disturbances encountered in nature. Bumblebees are expert commanders of the aerial domain as they persistently forage within complex terrain elements. Physical obstacles such as flowers produce local effects representative of a typified gust which threatens the precise control of intricate maneuvers. By tracking the three-dimensional dynamics of bees flying through gusts, we determined the sequences of motion that permit flight in three disturbance conditions: sideward, upward and downward gusts. Bees repetitively executed a series of passive impulsive maneuvers followed by active recovery maneuvers. Impulsive motion was unique in each gust direction, maintaining control purely by passive manipulation of the body. Bees pitched up and slowed-down at the beginning of recovery in every disturbance, followed by corrective maneuvers which brought attitudes back to their original state. Bees were displaced the most by the sideward gust, displaying large lateral translations and roll deviations. Upward gusts were easier for bees to fly through, causing only minor flight changes and minimal recovery times. Downward gusts severely impaired the control response of bees, inflicting strong adverse forces which sharply upset trajectories. Bees used interesting control strategies when flying in each disturbance, offering new insights into insect-scale flapping flight and bio-inspired robotic systems.

**Introduction**

Insects display a remarkable ability to engage in nimble control over their trajectory and attitude during flight. These flight characteristics have inspired great scientific endeavors into flapping wings, leading to knowledge of valuable concepts for flight such as unsteady lift mechanisms (Ellington et al., 1996; Dickinson et al., 1999; Sane, 2003), control capabilities (Sane and Dickinson, 2001; Deng et al., 2006) and underlying wing functions (Wootton, 1992; Usherwood and Ellington., 2002; Zhao et al., 2010). Studies on specific features of flying organisms have enabled intelligent inspiration in robotic design (Nakata et al., 2011; Ma et al., 2013; for review, see Shyy et al., 2016). However, most studies treat the flow environment as smooth, a major divergence from its true unsteady form in almost all flying scenarios.

The unpredictable conditions of the lower level of the atmosphere are ubiquitous. For those flying systems smaller in scale, the complex arena contains airflow that is highly changeable in strength and structure. Even away from local effects such as wakes of structures and vegetation the wind is highly turbulent (Watkins et.al., 2010). All flying animals use flapping wings rather than fixed wings to produce the aerodynamic forces necessary for flight. A number of studies allude to flapping in overcoming the effects of atmospheric conditions. Flapping wings have been shown to offer control and performance advantages in turbulence (Fisher, 2013), in vortices (Ortega-Jimenez, 2013; Ravi et al., 2015) and in gusts (Fisher, 2016; Viswanath and Tafti, 2010) relative to fixed or rotary wings at Reynolds numbers representative of insects. But the effects of these conditions on the dynamics of flapping flyers and the essential flight control behaviors which may assist in dealing with them are scarce.

Gusts and other atmospheric fluid structures are frequently referred to in the literature as damaging conditions which impede the control performance of small aerial systems (Watkins et al., 2006; Ravi et al., 2015. While turbulence and structured flow vortical streets caused by object wakes are significant on the broad scale, discrete gusts could be a critical element of the unsteady local flight aerodynamic condition at smaller scales and thus are particularly relevant to insects. In cluttered environments, studies declare the existence of vortical wakes, particularly Von Karman Streets that face insects as severe control challenges (Ravi, 2013; Ortega-Jimenez, 2013). The approximate scale of these vortices relevant to insects can be on the order of a few centimeters, matching the comparable scale of many insect wings. Relative to a flying insect navigating through one of these wakes, the adverse flow that interacts with the wings could be adequately described by a discrete gust containing local flow that meets the airborne surfaces predominantly in a singular direction.

Atmospheric airflow near rough terrain at low altitude (below 20m) has been shown to contain lateral turbulence intensities that increase closer to the surface and exceed vertical turbulence intensities which asymptotes to zero closer to the ground (Flay et al., 1978; Watkins et. al., 2006). On the contrary, atmospheric conditions in a cluttered environment have been shown to contain large (relative to the wingspan of a common insect) regions of vertical and lateral flow velocities (Ennos, 1999; Cresswell, 2010). That is, air flow relative to the scale of insects which can travel in any particular direction. These conditions representing gusts are created by the wakes of organic structures such as flower stems, leaves and other natural or artificial obstacles with wind flowing around them. For a small insect scoping out its approach to a flower for landing or feeding, discrete gusts could be of particular harm to their controllability.

A foraging mission for an insect on a typical day will involve sudden transitioning from regions of varying air states, flying across treacherous winds to a source of food or pollen, narrowing in on small and often dynamic landing sites, gleaning nectar on an often unstable platform and then navigating back for the return trip. The most aerodynamically challenging of these events is likely to be step changes between air scenarios and pinpointing a landing whereby complex flight maneuvers are impeded by unpredictable airflows. Approaching any solid object will involve traversing across shear layers near the surface, where the depth of the shear layer will likely be close to a few characteristic insect-wing dimensions. Local wakes including vortex shedding from surrounding vegetation could impinge on the insect from any orientation. The flow field is thus dominated by short scale changes from the wakes and vortices shed from plant structures.

The magnitude of the atmospheric wind varies with elevation, terrain and climatic conditions. It can vary from zero on days of calm (typically 5% of the time for non-cyclonic areas) to extreme, typified by the one-hundred-year return wind speed. It has been shown that in the last few meters from the Earth's surface the most likely speed is 3 m s$^{-1}$ and that for 95% of the time the speed is less than 10 m/s (for details see Watkins et. al. 1995). In this study, a single gust speed of 5 m s$^{-1}$ serves as a relevant basis for examining flying insects.

Recent studies demonstrate that organized body orientation maneuvers interlace the translatory motions observed among many flying insects in turning flight (Ristroph, 2012; Wang et al., 2003; Zeyghami and Dong, 2015), landing/take-off (Bode-Oke et al., 2017; Evangelista, 2010) and hovering flight (Sun, 2014). Attitude manipulation for control has also been found in involuntary (disturbance negotiation) flight scenarios in turbulence (Combes and Dudley, 2009) and vortices (Ravi et al., 2013; Ortega-Jiminez et al., 2013). However, our understanding of the effects and interactions that gusts play on the dynamics of insect-scale flapping flyers is shallow. The attitude adjustments that are a prerequisite for control of translatory motions in common flying environments could reveal important information pertaining to control for insect-scale flight.

We explored the flight dynamics of flying insects in strong gusts from three orthogonal directions. The sequences of trajectory and attitude changes in six degrees of freedom were tracked to measure the complete dynamic response of insect bodies. We recorded displacements and rates of change to quantify motion as well as acceleration to infer forces. Statistically significant encounters were gathered from the data and analyzed to build on our understanding of the effect that gusts have on centimeter-scale flapping wings.

**Materials and method**

**Experiment Setup**

Bumblebees (*Bombus ignitus*) sourced from a commercial breeder (Koppert; distributed by Arysta LifeScience; product name: Mini Polblack) were sustained in laboratory conditions. A foraging chamber of dimensions 1 m x 1 m x 0.8 m was accessible to the bees via a stagnant-air tunnel of length 1 m constructed from clear Perspex (Fig. 1A). The constant rectangular tunnel was designed with a cross-sectional dimension of 0.3 m x 0.3 m - sufficient space for aerobatic maneuvers and the application of strong in-flight disturbances that are relevant to normal outdoor conditions. In the foraging area, an array of artificial linalool-scented nectar flowers was provided for the bees to feed. Authentic flower pollen was also provided for collection adjacent the artificial flowers to ensure natural sustainment of the hive.

Following a several-day habituation period, 10 healthy foraging bees were captured and cold-anesthetized. Markers (Fig. 2A) were then affixed to each of the bees according to the method described in (Ravi et al., 2013). Marked bees were later released back into the foraging area where they were able to fully recover and resume regular transit between the foraging area and the hive.

Flights through the Perspex tunnel were perturbed by a strong wind gust in the form of a thin, high-velocity air sheet which acted in the middle of the tunnel length. Gusts were directed along the cross-section spanning perpendicular to the longitudinal axis of the tunnel and were operated continuously during experimentation (see Movie 1 for visual representation of setup). Airflow resembling a gust (a net flow of air moving in a particular direction) was produced using a plastic air-knife nozzle driven by a Teral VFZ 1000W ring blower with a restricted inlet.

The nozzle outlet was designed such that it could be inserted flush with the inner surface of any of the four walls comprising the tunnel (Fig. 1B), thereby enabling changeable airflow direction in any of the 90 deg positions without affecting the structure of the gust. On the face of the tunnel opposite to the gust, a meshed gap was created to allow air to escape and prevent flow recirculation. The strength of the gust was adjusted to contain an average flow velocity that applied a high-impulse force to flying bees without causing surface contact or loss of control leading to crash.

**Flow field characterization**

The gust was quantified using particle image velocimetry (PIV) calculated from imagery produced by a Photron high speed camera. With the gust in position, a pulsed laser sheet was projected along the middle of the tunnel such that a horizontal cross-section of the tunnel was illuminated. PIV measurements were conducted in multiple planar locations on the vertical axis spanning the total height of the gust (see Movie 1). Particles from evaporated olive oil were used to seed the air comprising the flow of the gust.

Gusts penetrated the stagnant air in the flight tunnel/interrogation volume at a mean velocity of approximately 4 m s$^{-1}$ (Fig. 3A). The mean velocity profile extending vertically up the tunnel cross-section varied minimally, indicating that the gust was reasonably consistent up the height of the tunnel. Maximum jet intensity near the outlet (approximately 20 mm from the tip of the outlet) was approximately 2.8 m s$^{-1}$ greater than the mean flow elsewhere in the gust. Average gust boundaries took a constant linear form from the outlet point to the exit aperture on the opposite wall. The gust sheet tapered out by 8°, causing a linear change in the horizontal width of the gust. On average, the gust width grew from 3 mm to 45 mm across the width of the tunnel. Shear layer fluctuations occurred at a high frequency (at least 60 Hz) and the width of these unsteady movements varied by up to 6 mm (at the center axis of the tunnel). Relative to the rate at which bees travelled through the tunnel, the shear layer fluctuations buffeted bees rapidly (at least 10 times during their traversal) and thus we consider the effects of spatial and temporal implications caused by the shear layer on the dynamic response negligible. To know precisely when bees penetrated the gust sheet, gust boundaries were defined by the average edge of the shear layer fluctuations across the entire width of the tunnel. The observed average fast forward flight velocity of bees was 0.44 m/s and the average body length was 21 mm. Thus, error in the exact location of gust penetration caused by shear layer fluctuations was in the order of 2-3 millimeters (1/10th of the average bee body length). Gusts positioned in three different 90deg positions were each identical in form and the flow exiting at the opposite aperture in the wall did not recirculate or affect the structure of the gusts in the calibrated region of any of the three directions.

**Flight dynamics analysis**

Bees were filmed using three Photron high-speed cameras recording at 2000 fps at a shutter speed of 1/5000s. Recordings were captured manually using a remote trigger as marked bees flew through the gust. The interrogation volume was defined by a rectangular prism of dimensions (160mm x 90mm x 100mm)(Fig. 2B) which served as a calibration frame for direct linear transformation (DLT). The entire free-flight dynamic response of bees to gusts was encapsulated within the interrogation volume.

Spatial data of bees were extracted from the recordings using DLTdv5; an open-source MATLAB-executed software which tracks positional information of markers via DLT (Hedrick, 2008). Each of the three points on the markers were tracked throughout the recorded duration of the respective flight. These data were then smoothed using a second order low-pass Butterworth filter with a cut-off frequency of 30 Hz. Translational and rotational rates and accelerations were calculated by performing numerical differentiation on the filtered positional data.

To determine the time at which bees began to enter the region of air impinged by the gust, we referred to our PIV interpretation of the structure and form of the gust. Drawing from this, it was estimated that the shear-layer fluctuated negligibly and that linear gust boundaries were clearly defined. This allowed us to produce a static 3D reconstruction of the gust in the same coordinate system used to measure and track bees. Thus, at each instance throughout the recorded flight of each bee, we were able to use this model to determine the gust position relative to the bee.

To gauge the amount of error existing between our deduced prediction of the gust location and the absolute size and magnitude of its actual shear layer fluctuations, we compared our spatial calculations of the gust to the dynamic information extracted from bees. Accounting for body length and determining the locations at which accelerations in the direction of the gust spiked, we were able to see where bees began to be perturbed by the gust. We compared both of these methods of determining gust location and confirmed that discrepancies were within an order of millimeters ($\mu=2.75mm$).

Other sources of error were inherently drawn from the delicate method of tracking dynamics from synchronized and calibrated cameras. We calibrated the three cameras at the beginning and end of each set of flight recordings to minimize the likelihood of inaccurate DLT transformations caused by changes in the camera positions. However, total removal of any chance of accidental bumps, shakes and knocks to the cameras was impractical. To measure and control this digitization error we limited the DLT error residual to a maximum value of 1 pixel (1 pixel was usually about 0.2mm corresponding to about 1/100 bee lengths). This also accounted for digitization error in the process of marker tracking which was limited by the number of pixels present in the recorded images. Distant marker perspectives in frames comprised of a smaller number of pixels (due to a low-flying bee), were tracked away from the actual marker centroid. For these frames, tracking was achieved manually on a frame-by-frame basis. This limited the contribution of inaccuracy due to residual error, allowing us to control the degree of error in the setup.

**Results**

**Flight phases and attitude maneuvers**
Bees negotiated gusts uniquely when flying through each of the three gust directions. All bees prompted a combination of impulsive attitude maneuvers—those caused involuntarily by the sheer force of the gust—followed by a series of recovery attitude maneuvers—those commanded voluntarily in resistance to the gust in pursuit of recovery. To determine the time lengths in each of these two phases, the moment of gust entry was taken as the start of the impulsive phase. We then computed an attitude acceleration curve to know the exact time bees initiated the recovery phase. We termed these two distinct chapters of flights the 'impulsive phase' and the 'recovery phase' respectively. Attitude maneuvers were unique in each gust direction, yielding interesting roll, pitch and yaw signatures that highlight potential strategies for governing control.

Bees flew in the center of the tunnel with a nominally level and stable body on approach to each gust. First sign of a bee entering a gust was usually visible by a sharp deflection in the antennae. Bees actively responded to the abrupt change in flow conditions approximately 22 ms later by displaying varied wing sweep range, rearwards shift of the mid-stroke wing position and extension of limbs. At this point, bees were already well into the flow disturbance, following through with a caused series of impulsive motions.

Attitude maneuvers in the impulsive phase of all gust directions were always in the direction of 'push' caused by the gust. Recovery maneuvers usually opposed the force of the gust and involved corrective efforts to regain stability towards the original trajectory. Three impulsive maneuvers were detected as bees flew through the sideward gust—one in each of roll, pitch and yaw. Bees rolled a mean magnitude of 29 deg left ($\mu$ = -28.6 ± 10.9 deg) immediately upon piercing the shear layer of the gust (Fig 5A). This was followed by an average clockwise yaw maneuver of 20 deg ($\mu$ = -20.4 ± 5.9 deg) away from the gust. Body acceleration in the direction of the gust only occurred after the first impulsive roll maneuver.

Bees flying into upward gusts exhibited smaller kinematic disruption than those flying into sideward gusts. A single impulsive pitch-up maneuver of around 7 degrees ($\mu$ = 7.1 ± 3 deg) occurred during the impulsive phase as bees penetrated the upward flow (Fig. 5B). This was coupled with general rotations in roll and yaw with no distinct pattern leading in to recovery. In the recovery phase, bees pitched back down beyond the original neutral position. This was probably to direct lift forces forward and produce forces for increasing forward flight speed by way of the 'helicopter' mode of control. The pitch-down maneuver occurred alongside corrective rotations in roll and yaw that oscillated between ±20 degrees throughout the rest of recovery.

The response of bees flying through downward gusts was more erratic than that of the upward gust. A single sharp pitch-down maneuver of approximately 14 deg ($\mu$ = 14.3 ± 6.80 deg) was observed during the impulsive phase when bees began to intercept the downward flow of air (Fig. 5C). During the gust-induced fall, bees extended their legs and produced a sharp pitch-up maneuver of about 47 deg ($\mu$ = 47.1 ± 4.10 deg) to mark the beginning of the recovery phase. Bees took a pitch down maneuver back to neutral during the recovery phase which was interlaced with large roll (±40 deg) and yaw (±11 deg) corrections. By the end of recovery, bees had regained steady attitudes but trajectories usually remained displaced.

Attitude deviation was greatest around the roll axis in all gust directions ($R_{side}$–$P_{side}$, P<0.0001; $R_{side}$–$Y_{side}$, P=0.0024; $R_{up}$–$P_{up}$, P<0.0007; $R_{up}$–$Y_{up}$, P=0.335; $R_{down}$–$P_{down}$, P<0.246; $R_{down}$–$Y_{down}$, P=0.086) (Fig. 6B). The horizontal gust (sideward) primarily engaged bees about roll and yaw whereas the vertical gusts (upward, downward) engaged bees about pitch due to orthogonally directed aerodynamic forces between gust directions. A prominent difference arose between the impulsive pitch maneuvers in the two vertical gusts. With the only difference between these two cases being the gust force direction, we found pitch to impulsively deviate roughly 10 deg more in the downward gust (pitch down) relative to the upward gust (pitch up)($P_{down}$–$P_{up}$, P=0.0208)(Fig. 6B). In sideward gusts, pitch deviated similarly, exceeding the upward gust by only a few degrees ($P_{side}$–$P_{up}$, P=0.304). Yaw attitudes were significantly greater in sideward gusts compared to those of both the upward and downwards gusts ($Y_{side}$–$Y_{up}$, P=0.0118; $Y_{side}$–$Y_{down}$, P<0.0001). Yaw motions in the vertical gusts were a result of recovery control methods as opposed to dealing with a direct yaw perturbation as with the horizontal gust.

While the sequence of attitudes in the impulsive phase was governed by the gust direction (i.e. all maneuvers in the push direction of the gust), we found that bees showed a pattern for regaining control in the recovery phase of flight through all gust directions. All bees pitched up in response to the disturbance at the start of recovery regardless of gust direction; this was followed by oscillatory corrective adjustments to roll and yaw whose magnitude varied between gusts and appeared to assist in the body motion.

**Body Kinematics**

Sideward gusts perturbed the lateral position of bees by a mean amount of 48mm ($\mu$ = 48.3 ± 3.2mm), while bees in upward and downward gusts were displaced vertically by 32 mm (32.1 ± 2.3mm) and 57 mm ($\mu$ = 56.9 ± 5.7 mm) respectively (Fig. 6A). Individual maneuvers during the impulsive phase were followed by matching trajectories. For example, the impulsive leftward roll/yaw maneuver was immediately followed by leftward motion in the lateral direction and the impulsive upward/downward pitch maneuvers were followed by motion in the vertical axis. The downward gust forced bees into a very rapid nosedive which caused a loss of altitude that exceeded the translational deviations of both upward and sideward gusts ($Z_{down}$–$Z_{up}$, P=0.0017; $Z_{down}$–$Z_{side}$, P<0.0013) (Fig. 6A). This altitude change was around three times greater than the altitude change in its rival vertical gust (upward). In the axes orthogonal to sideward gusts, bees lost an average of 5mm ($\mu$ = 5.36 ± 7.61mm) altitude during the sideward disturbance. The horizontal deviations in the vertical gusts were 36mm ($\mu$ = 35.9mm ± 2.89mm) in the upward gust and 31mm ($\mu$ = 31mm ± 2.27mm) in the downward gust—comparable to the magnitude of vertical deviation in the same gusts. Bees took immediate efforts to return to the centre of the tunnel in the sideward gust, whereas bees didn't seek to rapidly recover their altitude in the vertical gusts (Fig. 4).

In all three gust directions, attitude changes occurred most rapidly around the roll axis for all flights (Fig. 7B). Sideward gusts caused bees to roll at a mean rate of 2300 deg s$^{-1}$ ($\mu$ = 2311 ± 254 deg s$^{-1}$) and yaw at a mean of 1100 deg s$^{-1}$ ($\mu$ =1092 ± 82 deg s$^{-1}$). Upward gusts caused pitch to change at an average of 1000 deg s$^{-1}$ ($\mu$ = 1025 ± 89 deg s$^{-1}$) and downward gusts caused an angular rate at an average value of 1100 deg s$^{-1}$ ($\mu$ = 1058 ± 93 deg s$^{-1}$), nearly half of the gust-induced roll rates recorded in the sideward gust ($\dot{R}_{side}$–$\dot{P}_{up}$, P=0.0007; $\dot{R}_{side}$–$\dot{P}_{down}$, P=0.0016). Roll rates in the upward and downward disturbances occurred at mean magnitudes of 1400 deg s$^{-1}$ ($\mu$ = 1514 ± 133 deg s$^{-1}$) and 1700 deg s$^{-1}$ ($\mu$ = 1684 ± 209 deg s$^{-1}$) respectively. The roll rate reflected the resultant severity of the demanded response, indicating that upward gusts may be easier to fly through (Fig. 7B).

All bees slowed down after encountering a gust. Bees in sideward gusts slowed to approximately 0.13 m s$^{-1}$ ($\mu$ = 0.13 ± 0.04 m s$^{-1}$) while downward gusts slowed to a speed of around 0.07 m s$^{-1}$ ($\mu$ = 0.07 ± 0.02 m s$^{-1}$) (Fig. 7A). This value is significantly less than that of the upward gust, which slowed to a mean velocity of 0.17 m s$^{-1}$ ($\mu$ = 0.17 ± 0.03 m s$^{-1}$). Bees slowed down in accordance with the magnitude of disruption demanded by each gust. Lateral velocities peaked at about 0.5 m s$^{-1}$ ($\mu$ = 0.45 ± 0.03 m s$^{-1}$) in sideward gusts, far exceeding the longitudinal velocities in the disturbance. The upward velocity in upward gusts was comparable to the lateral velocities and averaged at around 0.2 m s$^{-1}$ ($\mu$ = 0.22 ± 0.03 m s$^{-1}$), marking only subtle changes in velocity between the three axes. The downward gust produced the greatest gust-induced velocity, seeing a downward velocity of approximately 0.5 m s$^{-1}$ ($\mu$ = 0.48 ± 0.04 m s$^{-1}$)—comparable to the lateral velocities caused in the sideward gust ($\dot{Z}_{down}$–$\dot{Y}_{side}$, P=0.93) but far greater than the upward velocities triggered by the upward gust ($\dot{Z}_{down}$–$\dot{Z}_{up}$, P<0.0001) (Fig. 7A).

After arresting the effects of a gust, bees resumed traversal through the tunnel. Bees expended a mean time of 0.18 s ($\mu$ = 0.18 ± 0.04 s) in transiting through the sideward gust (Fig. 9A)—from the moment of entry to a location 20mm beyond the other side of the gust sheet. Time required to negotiate the gust differed greatly for the upward gust, requiring approximately only 0.12 s ($\mu$ = 0.12 ± 0.05 s; $t_{side}$–$t_{up}$, P=0.033) to fly through, but compared similarly with the downward gust which on average consumed 0.21 s ($\mu$ = 0.21 ± 0.03 s; $t_{side}$–$t_{down}$, P=0.78). These total transit times form the sum of impulsive and recovery times taken in the phases of each flight. The impulsive phases of flights through all gust directions were approximately equal in all cases at about 0.06 s ($t_{i\,side}$–$t_{i\,up}$, P=0.91; $t_{i\,side}$–$t_{i\,down}$, P=0.52; $t_{i\,up}$–$t_{i\,down}$, P=0.67) (Fig. 9B). Bees took the longest time to recover in the downward gust ($\mu$ = 0.125 ± 0.03 s) in a similar interval to sideward gusts ($\mu = 0.121 \pm 0.04$ s). The discrepancy with its upward rival ($\mu = 0.051 \pm 0.05$ s; $t_{i\text{up}}-t_{i\text{down}}$, P=0.001) exceeded a factor of two.

We plotted a regression line between body attitude and acceleration in the horizontal plane to assess this relationship (Fig. 8). The impulsive roll maneuver in sideward gusts correlated strongly negative with lateral acceleration by 0.5 (Fig. 8A) purely by the passive 'sailboat' model. Recovery attitude maneuvers correlated moderately with lateral acceleration by 0.44. Likewise in vertical gusts, we found a moderate positive correlation between pitch maneuvers and longitudinal acceleration 0.55 in upward gusts (Fig. 8B). In downward gusts, the more violent pitch maneuvers correlated strongly with longitudinal acceleration 0.61 meaning that bees had to deal with the additional acceleration produced in the impulsive phase (Fig. 8C). The slope of the regression line in all cases could be expected to be about the magnitude of gravity (9.8m/s^2/rad) if altitudes were held constant. However, since altitudes were not stable, we calculated a significantly smaller slope of the regression line in the downward gust (4.9) compared to the upward gust (6.4) which outlines the interesting disparity between these two gusts.

## Discussion
### The effect of gusts on flight dynamics
The results reveal that bees tackle gusts from different directions with varying levels of difficulty. Sideward and downward gusts are significantly more difficult for bees to fly through than upward gusts—shown by lesser magnitudes of all dynamic parameters in upward gusts. Differences in the response from each gust direction arose from the recovery period of flight. The large horizontal deviations in the vertical gusts (comparable to the gust-direction deviations) and the minimal vertical deviations in the sideward gust shows that bees used the lateral plane for control adjustments more than the vertical plane. Impulsive maneuvers were endured for an equivalent amount of time across each of the gust directions (Fig. 9B). The comparable impulsive phase durations here signify that the differences in response of bees to other gust directions are not due to directional sensory sensitivity. The recovery phase captured inconsistencies across gusts which reflect the difficulties with flying through certain flow directions as shown by the larger variances between time intervals. Energy expenditure is to some degree proportional to time in flight. We can thus infer that the approximate energetic cost of flight in each condition is significantly dissimilar across different gust flow directions. This shows that downward gusts issue a more damaging and energy-sapping challenge, significantly more so than upward gusts and comparable to sideward gusts, although in sideward gusts bees appeared to have a more patterned response strategy for dealing with the disturbance. The independent response measures taken in each gust, both passively and actively, yield interesting strategies for control (Fig. 4).

In the upward and downward gusts, no differences in the disturbance exist between the two cases other than the direction of the gust relative to gravity. We speculate that previously unidentified aerodynamic interactions with flapping wings could occur when gusts strike the surface from above or from below. Flow phenomena have been shown to uniquely interact with flapping wing aerodynamics, confirming that certain conditions can be detrimental to flapping flight (Engels et al., 2016; Kolomenskiy et al., 2016; Ravi et al., 2016; Crall et al, 2016). Based on our data we postulate that gusts directed downward may interact with flow structures such as the LEV and wake effects at low Reynolds numbers. We see a distinct difference in the gradients of the regression lines (Fig. 9B and Fig. 9C). This shows that the lift force vector was more severely impaired in the downward gust, likely a result of the destabilizing effect of the gust on flow structures which are critical to maintain the flight forces necessary for insect flight. Some studies have previously hinted that downward gusts can interfere with the LEV of insects (Jones and Yamaleev, 2016, Jones and Yamaleev, 2012). In upward gusts, these mechanisms could be somewhat protected by the higher-pressure surface below. Our results indicate the presence of unknown aerodynamic interactions with flapping wings, although a fundamental study is required to uncover the underlying processes by which these occur. We aim to measure the reason for the discrepancy in future work with a representative robotic flapper that could help to solve these unknowns.

In the case of the sideward gust, bees always yawed towards the direction of the gust rather than continuing forward-directed flight through the tunnel. It could be that bees are passively stable around the z-axis as observed in hawkmoths (Nguyen et al., 2016). Recordings showed that bees consistently shifted mean wing stroke angles rearwards upon reaction of the disturbance. This rearward shift could cause a stabilizing moment about the Z axis which assists the observed yaw motion (effectively converting the sideward gust into a frontal gust relative to the bee). Flapping with an average stroke angle behind the center of gravity of the body, bees can be expected to generate additional yaw torque as the wing collects forces from the gust at some distance behind the CG. This behavior likens to the inherent passive stability of the tail of a conventional passenger aircraft or a feathercock. In addition, limb extensions manipulate stability by augmenting aerodynamic force acting on the body and thereby amplifying the resultant stabilizing body torques. Limb extensions also shift the center of gravity downwards causing a more 'bottom-heavy' distribution of mass, which is a known attribute for stable control in flying insects (Liu et al., 2012).

The tendency of bees to move in correspondence with their attitude is a method for control that enables flapping flyers to produce flight forces for motion in the horizontal plane. The 'helicopter mode' has been witnessed amongst bumblebees (Ravi et al., 2016; Ravi et al., 2013), hawkmoths (Greeter et al., 2016) and fruit flies (Muijres, 2015) along with a number of bird species during routine flight maneuvers (Thomas and Taylor, 2001; Ros et al., 2011). This method of control and an idealized 'sailboat' model has been seen in bees struck by lateral flows (Ravi et al., 2016). Here, we show that the helicopter mode is employed throughout the sequence of maneuvers demanded by gust perturbations when the flow axis is directed sidewards, upwards and downwards. Bees used the helicopter mode of control during impulsive and recovery maneuvers when struck by gusts from all directions. When flying through steady air, bees used the helicopter mode to undergo side-to-side casting motions. This was also true in the impulsive phase of flight although bees did not have active control of their body motion due to limits to their reaction time. Rather, all resultant forces acting on the body were passively commanded by the interaction of the buffeting gust on the insect body; yet nonetheless in agreement with the helicopter control strategy.

In steady flight, bees varied longitudinal forward-flight velocity along the tunnel between 0.22 m s$^{-1}$ and 0.57 m s$^{-1}$. We found that gust entry velocity correlated with transit time by r=0.72 (sideward). This demonstrates that the velocity of bees when entering a gust is a responsible factor in the severity of the resultant disturbance. This is likely due to the association of velocity with disturbance impulse time and the resistance of inertial changes by gust forces. For both of these reasons, bees may benefit from barging through gusts rather than taking it slow in the case of a single discrete disturbance. However, a trade-off arises between sensory detection time and gust impulse time. Bees that travel faster would have to travel through more of the unsteady, potentially dangerous flow conditions before assessing the threat it poses. On the other hand, bees that travel slower would receive greater aerodynamic impulses from the gust. In our study, most bees took the former and immediately slowed down to favor sensory awareness. Those few that surged through the gust were able to deal with the recovery briskly but this may not be the case for imperfect natural conditions less discrete than that of this study.

**Concluding remarks**

This work studies a skillful natural flyer to identify several control-related behaviors during its flight through gusts. The results show that bumblebees tackle gusts from different directions in different ways, but always pitch up and slow down upon meeting each disturbance. Bees are shown to be more affected by downward gusts than upward gusts. Sideward gusts cause a disturbance similar in magnitude to downward gusts, though bees appear to have a robust method for overcoming the more common lateral hindrance. Bees yawed into the sideward gust, be it passively or actively, to increase the frontal component of flow, thereby favoring aerodynamic force production and control. These control strategies are useful in uncovering the clever flight conducts of volant insects, while providing potentially useful bio-inspired ideas towards the development of similar scale flying robots.

**Acknowledgments**

The authors would like to thank Hiroto Tanaka and all members of Hao Liu Lab for assistance with equipment and useful input during experimentation. The authors also thank Mark Finnis for his valuable software and support during data analysis.

**Reference list**

**Bode-Oke A T, Zeyghami S., Dong H.** (2017). Aerodynamics and flow features of a damselfly in takeoff flight. *Bioinspir. Biomim.* **12,** 056006.

**Cheng, B. and Deng, X.** (2011). Translational and rotational damping of flapping flight and its dynamics and stability at hovering. *IEEE Trans. Rob*. **27**, 849-864.

**Combes, S. A. and Dudley, R.** (2009). Turbulence-driven instabilities limit insect flight performance. *Proc. Natl. Acad. Sci. USA* **106**, 9105-9108.

**Crall, J. D., Chang, J. J., Oppenheimer, R. L. and Combes, S. A.** (2016). Foraging in an unsteady world: bumblebee flight performance in field realistic turbulence. *Interface Focus* **7**, 20160086.

**Cresswell, J.E., Krick, J., Patrick, M.A., Lahoubi, M.** (2010). The aerodynamics and efficiency of wind pollination in grasses. *Funct. Ecol.* **24**, 706-713.

**Deng, X., Schenato, L. and Sadtry, S. S.** (2006). Flapping flight for biomimetic robotic insects. Part II, flight control design. *IEEE Trans. Rob.* **22**, 789 -803.

**Dickinson, M. H., Lehmann, F. O. and Sane, S. P.** (1999). Wing rotation and the aerodynamic basis of insect flight. *Science* **284**, 1954-1960.

**Ellington, C. P., Van den Berg, C., Willmott, A. P. and Thomas, A. L. R.** (1996). Leading-edge vortices in insect flight. *Nature* **384**, 626-630.

**Engels, T., Kolomenskiy, D., Schneider, K., Lehmann, F.-O., Sesterhenn, J.** (2016) Bumblebee flight in heavy turbulence. *Phys. Rev. Lett.* **116**, 028103.

**Ennos, A R.,** (1999). The aerodynamics and hydrodynamics of plants. *J. Exp. Biol.* **202**, 3281–3284.

**Evangelista, C., Kraft, P., Dacke, M., Reinhard, J. and Srinivasan, M. V.** (2010). The moment before touchdown: landing manoeuvres of the honeybee Apis mellifera. *J. Exp. Biol.* **213**, 262-270.

**Fisher A.** (2013). The effect of freestream turbulence on fixed and flapping micro air vehicle wings. *PhD Thesis*, RMIT University, Melbourne, Australia.

**Fisher, A., S. Ravi, S. Watkins, J. Watmuff, C. Wang, H. Liu, and P. Petersen.** (2016). "The gust-mitigating potential of flapping wings". *Bioinspir. Biomim.* **11**, 046010.

**Flay, R.G.J.** (1978). Structure of a Rural Atmospheric Boundary Layer near the Ground", *PhD Thesis*, University of Canterbury, New Zealand.

**Greeter, J. S. and Hedrick, T. L.** (2016). Direct lateral maneuvers in hawkmoths. *Biol. Open* **5**, 72-82.

**Hedrick, T. L.** (2008). Software techniques for two- and three-dimensional kinematic measurements of biological and biomimetic systems. *Bioinsp. Biomim.* **3**, 034001.

**Jones M., and Yamaleev N.** (2016). Effect of lateral, downward, and frontal gusts on flapping wing performance. *Computers & Fluids* **140,** 175-190.

**Jones M., and Yamaleev N.** (2012). The effect of a gust on the flapping wing performance. *AIAA*, 1080.

**Kolomenskiy, D., Ravi, S., Takabayashi, T., Ikeda, T., Ueyama, K., Engels, T., Fisher, A., Tanaka, H., Schneider, K., Sesterhenn, J., Liu, H.** (2016). Added costs of insect-scale flapping flight in unsteady airflows. *arXiv preprint,* arXiv:1610.09101.

**Liu, B., Ristroph, L., Weathers, A., Childress, S. and Zhang, J.** (2012). Intrinsic stability of a body hovering in an oscillating airflow. *Phys. Rev. Lett.* **108**, 068103.

**Ma, K. Y., Chirarattananon, P., Fuller, S. B. and Wood, R. J.** (2013). Controlled flight of a biologically inspired, insect-scale robot. *Science* **340**, 603-607.

**Muijres, F. T., Elzinga, M. J., Iwasaki, N. A. and Dickinson, M. H.** (2015). Body saccades of Drosophila consist of stereotyped banked turns. *J. Exp. Biol.* **218**, 864-875.

**Nakata, T., Liu, H., Tanaka, Y., Nishihashi, N., Wang, X. and Sato, A.** (2011). Aerodynamics of a bio-inspired flexible flapping-wing micro air vehicle. *Bioinspir. Biomim.* **6**, 045002.

**Nguyen, Anh Tuan, Han, Jong-Seob, & Han, Jae-Hung.** (2016). Effect of body aerodynamics on the dynamic flight stability of the hawkmoth Manduca sexta. *Bioinspir. Biomim.* **12**, 016007.

**Ortega-Jimenez, V. M., Greeter, J. S. M., Mittal, R. and Hedrick, T. L.** (2013). Hawkmoth flight stability in turbulent vortex streets. *J. Exp. Biol.* **216**, 4567-4579.

**Ravi, S., Crall, J. D., Fisher, A. and Combes, S. A.** (2013). Rolling with the flow: bumblebees flying in unsteady wakes. *J. Exp. Biol.* **216**, 4299-4309.

**Ravi, S., Crall, J. D., McNeilly, L., Gagliardi, S. F., Biewener, A. A. and Combes, S. A.** (2015). Hummingbird flight stability and control in freestream turbulent winds. *J. Exp. Biol.* **218**, 1444-1452.

**Ravi S, Kolomenskiy D, Engels T, Schneider K, Wang C, Sesterhenn J, Liu, H.** (2016). Bumblebees minimize control challenges by combining active and passive modes in unsteady winds. *Scientific Reports* **6**, 35043.

**Ristroph, L., Bergou, A. J., Ristroph, G., Coumes, K., Berman, G. J., Guckenheimer, J., Wang, Z. J. and Cohen, I.** (2010). Discovering the flight autostabilizer of fruit flies by inducing aerial stumbles. *Proc. Natl. Acad. Sci. USA* **107**, 4820-4824.

**Ristroph L, Bergou AJ, Berman GJ, Guckenheimer J, Wang ZJ, Cohen I.** (2012). Dynamics, control, and stabilization of turning flight in fruit flies. *Natural locomotion in fluids and on surfaces*, 83–99.

**Ristroph, L., Ristroph, G., Morozova, S., Bergou, A. J., Chang, S., Guckenheimer, J., Wang, Z. J., Cohen, I.** (2013). Active and passive stabilization of body pitch in insect flight. *J. R. Soc. Interface* **10**, 20130237.

**Ros, I. G., Bassman, L. C., Badger, M. A., Pierson, A. N. and Biewener, A. A.** (2011). Pigeons steer like helicopters and generate down- and upstroke lift during low speed turns. *Proc. Natl. Acad. Sci. USA* **108**, 19990-19995.

**Sane, S. P. and Dickinson, M. H.** (2001). The control of flight force by a flapping wing: lift and drag production. *J. Exp. Biol.* **204**, 2607 -2626.

**Shyy, W. Kang, C.K. Chirarattananon, P. Ravi, S. Liu, H.** (2016). Aerodynamics, sensing and control of insect-scale flapping-wing flight. *Proc. R. Soc. A Math. Phys. Eng. Sci.*, **472**.

**Sane, S. P.** (2003). The aerodynamics of insect flight. *J. Exp. Biol*. **206**, 4191-4208.

**Sun, M.** (2014). Insect flight dynamics: stability and control. Rev. Mod. Phys. **86**, 615-646.

**Thomas, A. L. R. and Taylor, G. K.** (2001). Animal flight dynamics. I. Stability in gliding fight. *J. Theor. Biol.* **212**,399-424.

**Usherwood, J. R. and Ellington, C. P.** (2002a). The aerodynamics of revolving wings – I. Model hawkmoth wings. *J. Exp. Biol.* **205**, 1547 -1564.

**Vance, J. T., Faruque, I. and Humbert, J. S.** (2013). Kinematic strategies for mitigating gust perturbations in insects. *Bioinspir. Biomim.* **8**, 016004.

**Viswanath K. and Tafti, D. K.** (2010). Effect of frontal gusts on forward flapping flight. *AIAA J.* **48**, 2049.

**Wang, H., Zeng, L., Liu, H. and Chunyong, Y.** (2003). Measuring wing kinematics, flight trajectory and body attitude during forward flight and turning maneuvers in dragonflies. *J. Exp. Biol.* **206**, 745-757.

**Watkins S., Saunders J.W., Hoffmann P.H.** (1995). Turbulence experienced by moving vehicles. Part I. Introduction and turbulence intensity. *J. Wind Eng. and Ind. Aero.* 57, 1-17.

**Watkins S., Thompson M., Loxton B., Abdulrahim M.** (2010). On Low Altitude Flight Through The Atmospheric Boundary Layer. *International Journal of Micro Air Vehicles*, **2,** 55-68.

**Watkins S., Milbank J., Loxton B. J. and Melbourne W. H.** (2006). Atmospheric Winds and their Effects on Micro Air Vehicles. *Journal of the American Institute of Aeronautics and Astronautics* **44**, 2591-2600.

**Wootton, R. J.** (1992). Functional morphology of insect wings. *Annu. Rev. Entomol.* **37**, 113 -140.

**Zeyghami S., Dong H.** (2015). Coupling of the wings and the body dynamics enhances damselfly maneuverability. *arXiv preprint*, arXiv:1502.06835.

**Zhao, L., Huang, Q., Deng, X. and Sane, S. P.** (2009). Aerodynamic effects of flexibility in flapping wings. *J. R. Soc. Interface* **7**, 485-497.

**Figures**

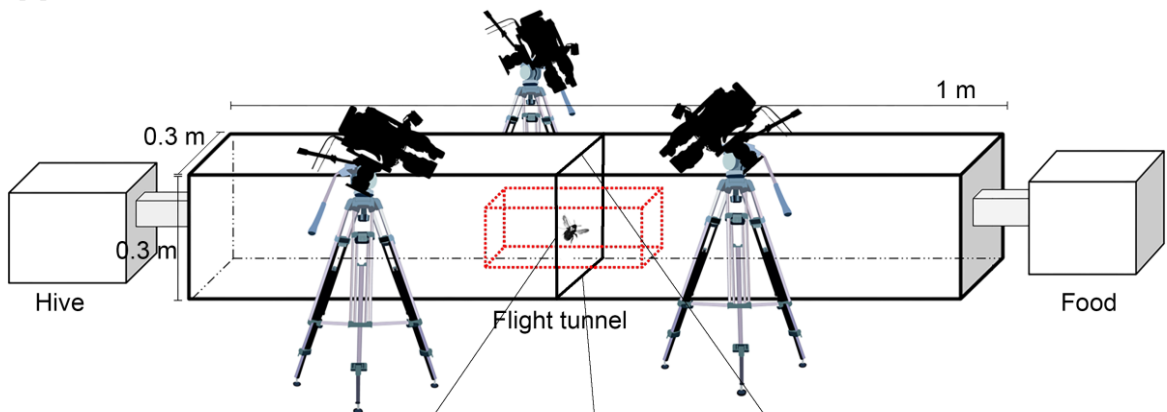

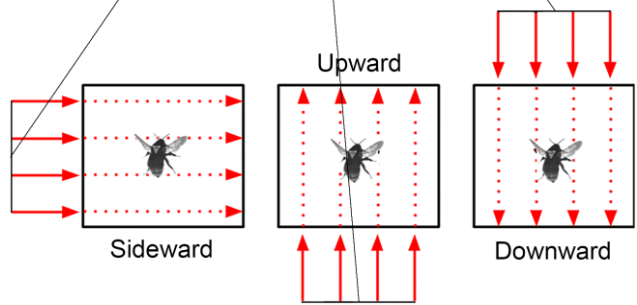

Fig. 1. (A) Setup of the bumblebee domain showing arrangement of the three interconnected sectors: the hive area, flight tunnel and feeding chamber. Configuration of recording equipment relative to the calibrated volume (represented by a red dotted prism around bee) shown in reference to the flight tunnel. (B) Cross-section views of the middle of the flight tunnel where identical gusts are directed sideward, upward and downward across this plane.

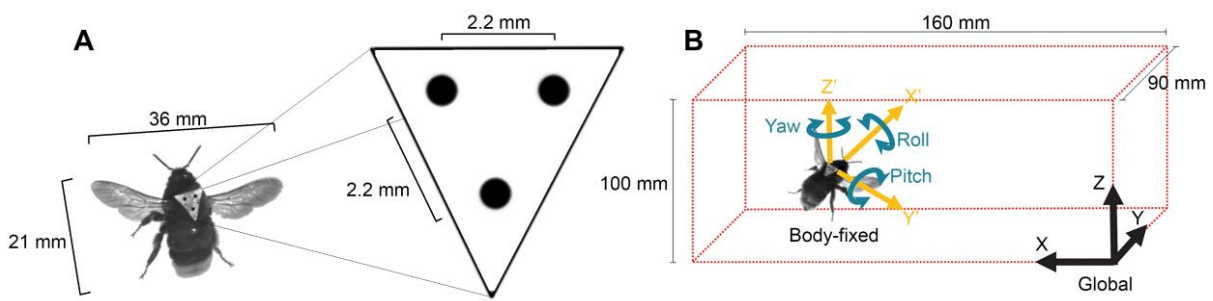

Fig 2. Overview of the arrangement for tracking bumblebee body dynamics. (A) Layout of triangular markers affixed to the thorax of bees. Markers were aligned with longitudinal axis of bees such that the long arm always faced rearwards. Bumblebee dimensions are averages based on DLT measurements taken during steady flight. (B) Global and body-fixed coordinate systems defined within the calibrated volume.

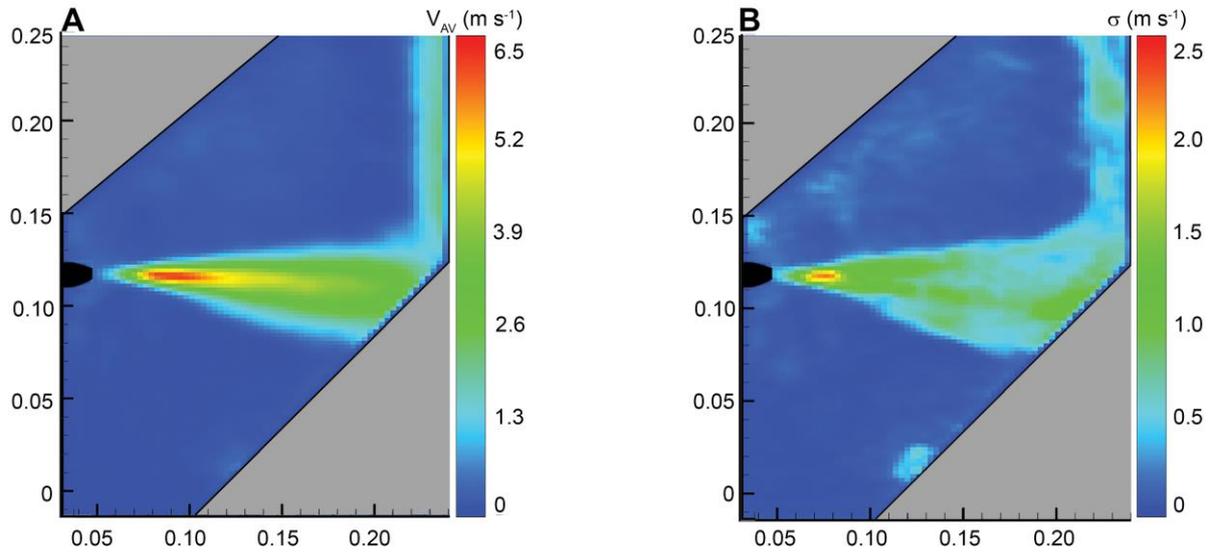

Fig. 3. Digital quanftification of the gust showing (A) average gust velocity calculated from 300 samples captured in one second and (B) standard deviation of the gust velocity. To capture gust in its actual operating state (in the flight tunnel), a laser sheet was projected diagonally into the cross-sectional field of interest in the gust midpoint. Hence, resultant images are cropped in a slanted format to capture the structure of the gust.

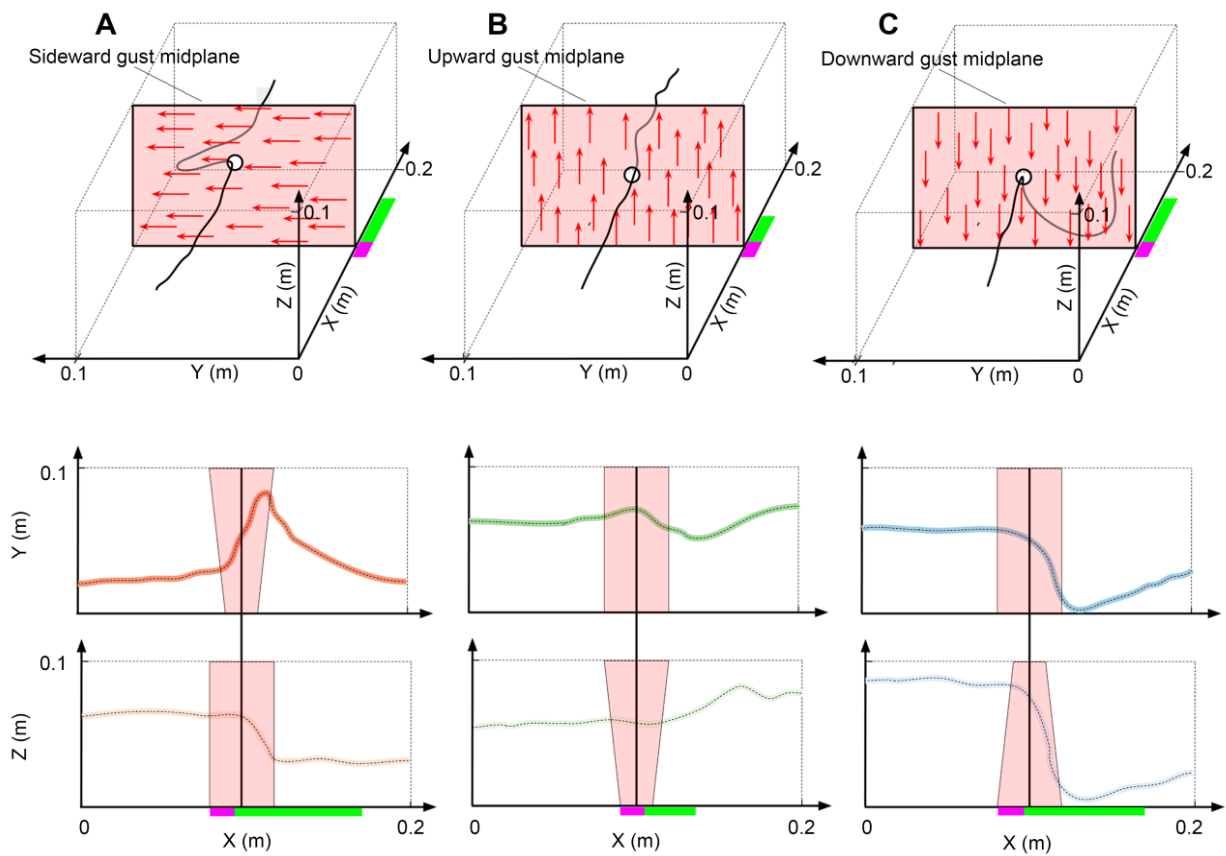

Fig. 4. Typical flight trajectories through the calibrated volume in: (A) a sideward gust (B) an upward gust (C) a downward gust. Side and top views of trajectory are displayed below in terms of Z and Y respectively.

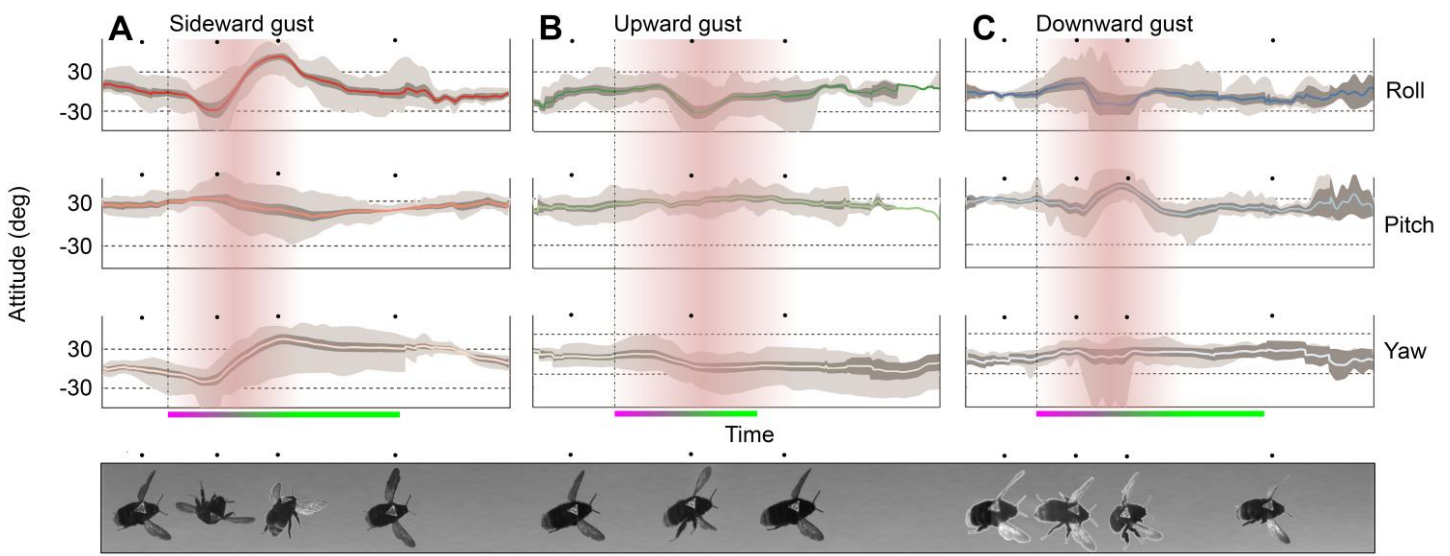

Fig. 5. Mean attitude maneuvers of bees flying through (A) sideward gusts (B) upward gusts and (C) downward gusts. Axes are plotted against time referenced to the time instance at which bees first enter the gust. Gusts are illustrated in red where fading color represents the time at bees were less likely to be within the gust. Dark shading represents the standard error of the mean attitude. Light shading shows the maximum and minimum angular displacements recorded. Response phases are shown in a purple (impulsive) zone and a green (recovery) zone. Corresponding image snapshots of bees during major maneuvers are represented sequentially below each set of plots (D).

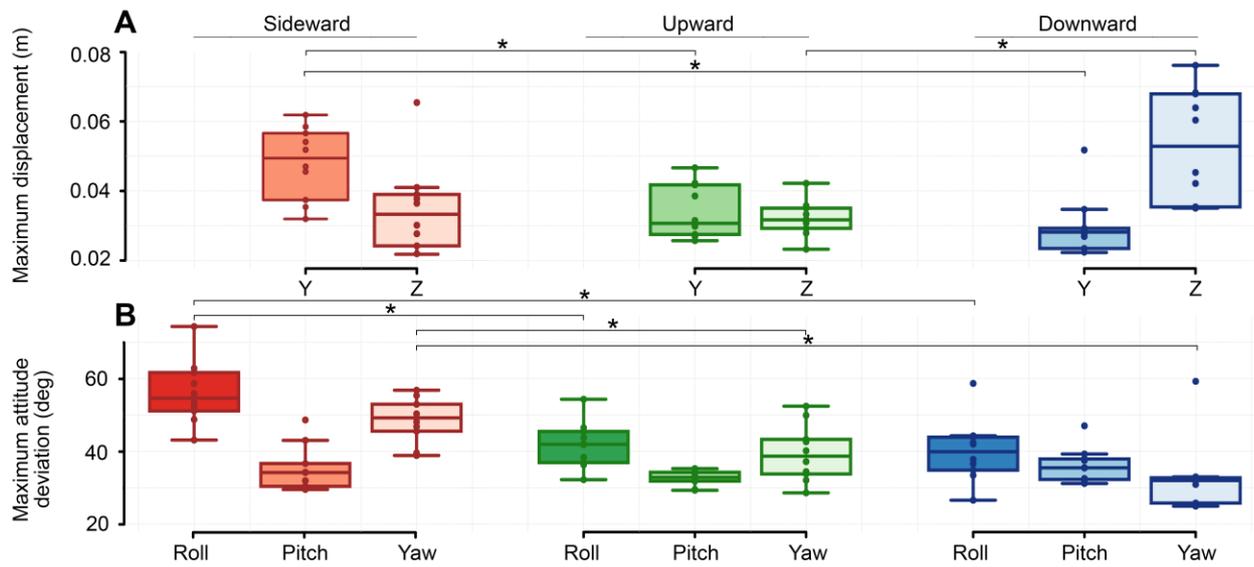

Fig. 6. Box plots of (A) maximum translation and (B) maximum rotation deviations for bees flying in three different gust directions. Significance is shown by asterisks where the p-value between two independent samples is less than 0.05.

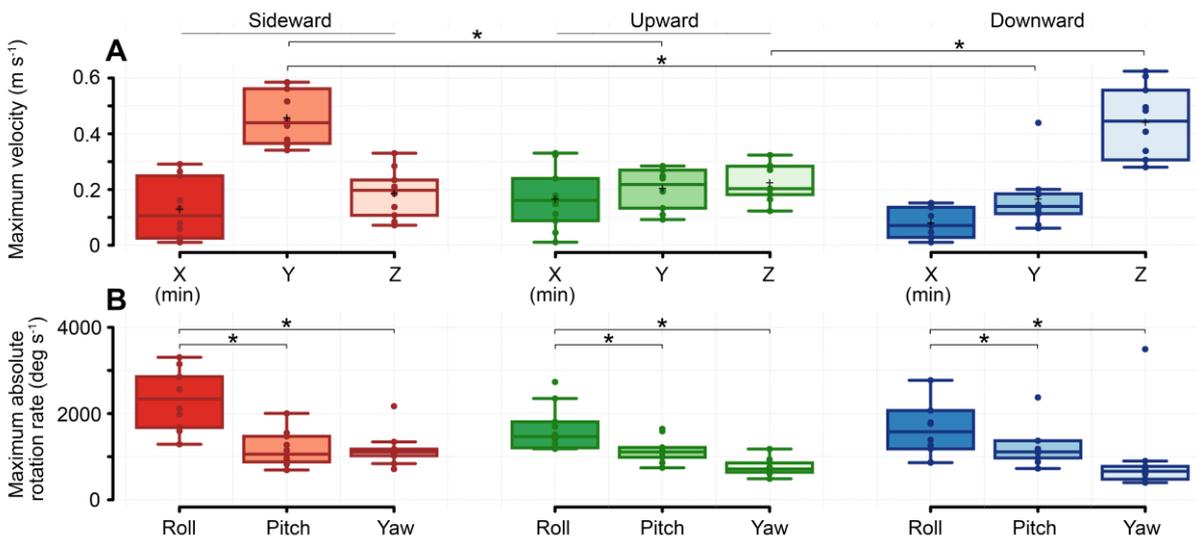

Fig. 7. Box plots of (A) maximum translation and (B) maximum rotation rates in three gust directions. Maximum values are taken as the 95th percentile of data to eliminate error arising from spikes in the numerically differentiated values.

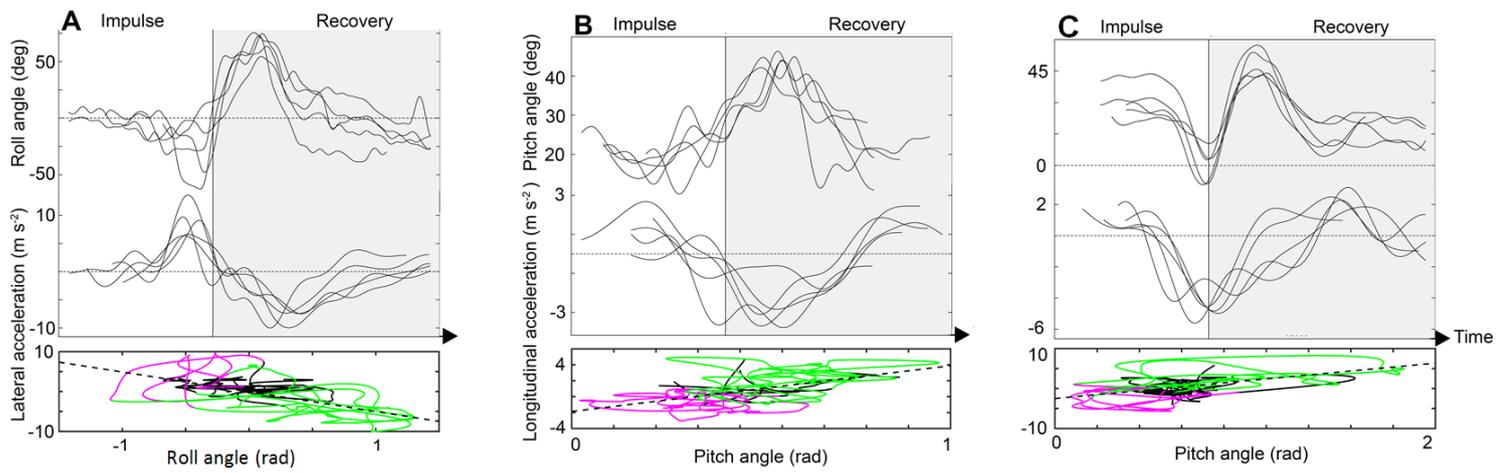

Fig. 8. Respective bee attitudes and accelerations in the horizontal plane distinguished by the impulsive phase (white background) and recovery phase (grey background) in (A) sideward gusts, (B) upward gusts and (C) downward gusts. Regression lines are plotted on the set axes below where flight phases are distinguished highlighted in purple (impulsive) and green (recovery).

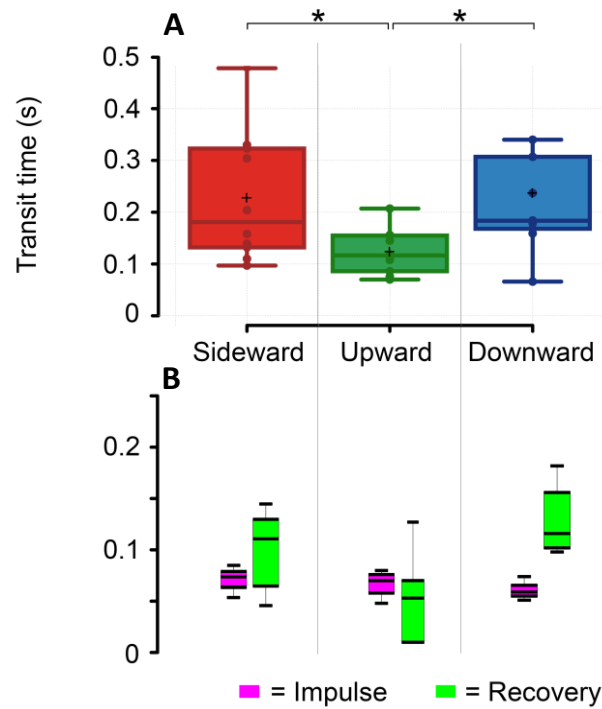

Fig. 9. (A) Box plots showing time taken to transit through the entire disturbance caused by the gust. (B) Box plots of time taken through the individual impulse (purple) and recovery (green) phases in each of the gust directions.